\documentclass[a4paper]{jpconf}
\usepackage{graphicx}
\usepackage{amsmath}
\usepackage{cite}
\usepackage{subfigure}

\begin{document}
\title{First 2D-ACAR Measurements on Cu with the new Spectrometer at TUM}

\author{J A Weber$^1$, P B\"oni$^1$, H Ceeh$^1$, M Leitner$^{2,3}$ and Ch Hugenschmidt$^{1,2}$}

\address{$^1$ Physik-Department E21, Technische Universit\"at M\"unchen, James-Franck-Stra\ss e 1, 
85748 Garching, Germany}
\address{$^2$ FRM II, Technische Universit\"at M\"unchen, Lichtenbergstra\ss e 1, 85748 Garching, Germany}
\address{$^3$ Physik-Department E13, Technische Universit\"at M\"unchen, James-Franck-Stra\ss e 1, 85748 Garching, Germany}

\ead{josef-andreas.weber@frm2.tum.de}

\begin{abstract}
The two-dimensional measurement of the angular correlation of the positron annihilation radiation (2D-ACAR) is a powerful tool to investigate the electronic structure of materials. Here we report on the first results obtained with the new 2D-ACAR spectrometer at the Technische Universit\"at M\"unchen (TUM).
To get experience in processing and interpreting 2D-ACAR data, first measurements were made on copper. The obtained data are treated with standard procedures and compared to theoretical calculations. It is shown that the measurements are in good agreement with the calculations and that the Fermi surface can be entirely reconstructed using three projections only.  
\end{abstract}

\section{Introduction}
When a positron annihilates in a solid with an electron the resulting annihilation radiation carries the momentum information of the electron-positron pair. This information is contained in a Doppler shift from the mean value of 511\,keV, and the angular deviation from exact antiparallel directions of the two $\gamma$ quanta. With 2D-ACAR the latter is measured, which yields a specific projection of the two $\gamma$ momentum density $\rho^{2\gamma}$. Since a positron thermalizes within a few ps \cite{RevModPhys.60.701} the angular deviation is caused mainly by the electron momentum in the solid. 

The low momenta that are measured with 2D-ACAR are due to occupied states near the Fermi surface in the case of a metal. In these states, any electron is in a delocalised Bloch state. If we represent the according wave function in the basis of momentum eigenfunctions, it has to be expressed by a sum over different momentum components. Mathematically $\rho^{2\gamma}$ is given by

\begin{equation}
\rho^{2\gamma}({\bf p})\propto \sum_{j,\mathbf{k}:E(j,\mathbf{k})<E_{\rm F}} \left| {\int e^{-i{\bf p \cdot r}}\Psi_{j,{\bf k}}^{\rm ep}({\bf r,r}){\mathrm d}{\bf r} }\right| ^2
\end{equation} 
where the sum goes over all the states ${\bf k}$ below the Fermi level in the $j$th band. The wave function of the positron and an electron in the band $j$ and state ${\bf k}$ is given by

\begin{equation}
\Psi^{\rm ep}_{j,{\bf k}}({\bf r,r'})=\Psi^{\rm e}_{j,{\bf k}}({\bf r})\Psi^{{\rm p}}({\bf r'})\sqrt{\gamma_{j,{\bf k}}({\bf r, r'})}
\end{equation}
with a general enhancement factor $\gamma_{j,{\bf k}}({\bf r, r'})$, which accounts for the interaction of the positron with the electron. If we assume an independent particle model, i.e. $\gamma_{j,{\bf k}}({\bf r, r'})=1$, and put in the Fourier expansion $\Psi^{\rm e}_{j,{\bf k}}({\bf r})\Psi^{{\rm p}}({\bf r})=\sum_{\rm G}C_{j,{\bf G}}\exp{\left(i({\bf k+G}){\bf r}\right)}$ with reciprocal lattice vectors $\bf G$ we get:

\begin{equation}
\label{eq:lcw_erweitert}
\rho^{2\gamma}(\mathbf{p})\propto \sum_{j,{\bf k}} \Theta(E_{\rm{F}}-E_{j,{\bf k}}) \sum_{\bf G} \left|{C^j_{\bf G}({\bf k})} \right| ^2 \delta_{\bf p-k,G}
\end{equation}
From this equation we see that filled bands give a continuous distribution and bands crossing the Fermi level yield breaks which are distributed through the reciprocal space. An even more descriptive result is achieved if we define the following function $n({\bf P})$  of a vector in reciprocal space $\bf P$, assuming a constant positron wave function:

\begin{equation}
\label{eq:lcw_reduziert}
n(\mathbf P)=\sum_{\mathbf G} \rho^{2\gamma}(\mathbf P+\mathbf G) \propto \sum_{j,\mathbf k} \Theta(E_{\mathrm F}-E_{ j,\mathbf k})\text{\;.}
\end{equation}
The operation described by equation (\ref{eq:lcw_reduziert}) is called the Lock-Crisp-West (LCW) procedure \cite{Lock1973}. With latter equation the $\rho^{2\gamma}$ of equation (\ref{eq:lcw_erweitert}) can be represented in the reduced zone scheme.
The LCW procedure is applicable for $\rho^{2\gamma}$ and its projections in the main symmetry directions, i.e. the measured 2D-ACAR spectra.

The method used here to calculate the full three dimensional $\rho^{2\gamma}$ is based on the maximum entropy algorithm (MEA) \cite{Gull1978}. Although a MEA to de-convolve ACAR spectra is known to have positive effects on reconstructing $\rho^{2\gamma}$ \cite{0295-5075-32-9-012}, it was just recently, that a MEA has successfully been applied for reconstructing the 3D momentum density from 2D-ACAR measurements  \cite{springerlink:10.1007/s00339-011-6350-z}. The idea behind the MEA can be found, e.g, in \cite{Burch1983113}. The main purpose of the reconstruction is to get a three dimensional density $D$  which agrees within the 2D-ACAR spectra $M_i^\alpha$ (where $i=1...N^2$ runs over all pixel) for all measured angles $\alpha$ with the statistical error $\sigma_i$.  It could be realized by minimizing the following expression, where $R^\alpha_i(D)$ is the Radon transformation, i.e. the projection, of the density $D$ at an angle of $\alpha$:

\begin{equation}
\chi^2=\sum_\alpha \sum_i \frac{\left(R^\alpha_i\left(D\right)-M_i^\alpha\right)^2}{\sigma_i^2}
\end{equation} 

If $M$ is of size of $N^2$, then $D$ has a size of $N^3$. Consequently, the minimization problem is highly under-determined. This is the reason why the MEA is employed. In order to solve the minimization an additional criterion is used: The resulting density $D$ should be flat. For this purpose, a Lagrange function with $\chi^2$ and an entropy function $\sum_j D_j \ln{\left( D_j\right)}$ is constructed by the use of the Lagrange multiplier $\lambda$:

\begin{equation}
 L(D)=\sum_j D_j \ln{\left( D_j\right)}-\lambda \chi^2
\end{equation}
$L$ has to be maximized in order to obtain $D$. The solution to this problem can be written in an iterative form \cite{Skilling1984}

\begin{equation}
D_i^{(n+1)}=A\frac{D_i^{(n)} \exp{\left(-\lambda\frac{\partial \chi^2}{\partial D_i}\right)}}{\sum_j D_j^{(n)} \exp{\left(-\lambda \frac{\partial \chi^2}{\partial D_j}\right)}}
\end{equation}
with the normalisation constant A.

In case of $\rho^{2\gamma}$, the quality of the reconstruction can be improved and the calculation time is reduced if the full symmetry of the reciprocal lattice is taken into account.  Doing this, the MEA becomes a fast and reliable method to recreate the three dimensional $\rho^{2\gamma}$.  

\section{Experiments}

Measurements were performed at the new 2D-ACAR spectrometer at the TUM in Garching. The details of the set-up are reported in detail by H. Ceeh et al. \cite{ceeh:043905}.
A single crystalline copper sample was prepared from commercially available copper single crystal by orienting and cutting. To remove lattice defects produced by cutting, the sample was electrolytically polished.

During the measurement the sample was cooled below 10$\,$K. A spectrum with more than $10^8$ counts was recorded for 
projections along the main symmetry axes [100], [110] and [111]. All data were symmetrised according to their point group in order to enhance statistics. 
For the simulation of 2D-ACAR spectra, the Munich SPR-KKR package, version 5.4 by H. Ebert et al. was used \cite{0034-4885-74-9-096501}. 

\section{Results and Discussion}

\begin{figure}
	\centering
	\subfigure[\label{fig:vergleich} Measured (left) and calculated (right) anisotropy of the 2D-ACAR spectrum.]{\includegraphics[width=0.37\textwidth]{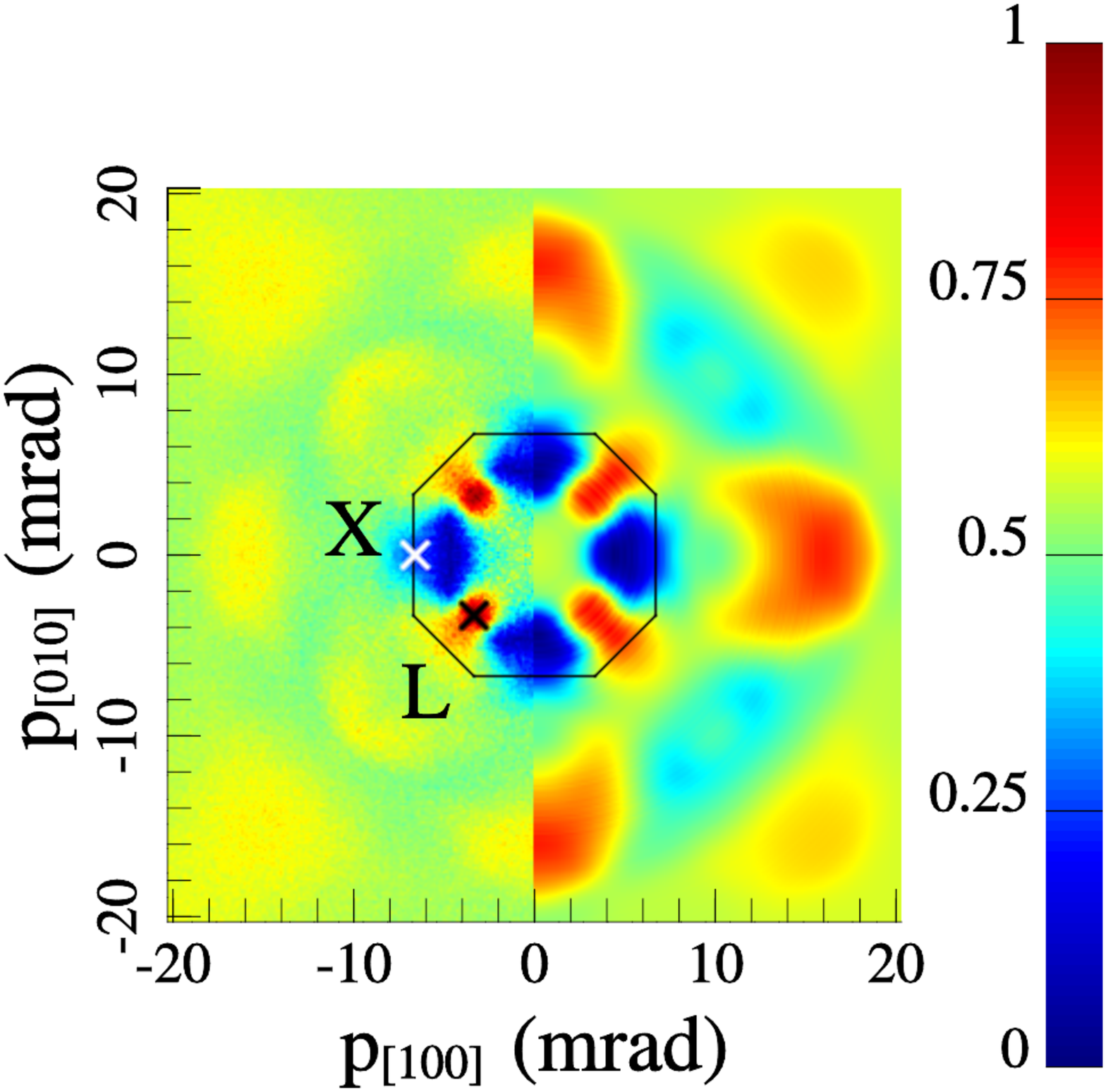}}\qquad
	\subfigure[\label{fig:vergleich2} LCW folded density in the {$[001]$} direction: measurement (left), calculation (right bottom) and calculation convolved with the experimental resolution (right top).]{\includegraphics[width=0.37\textwidth]{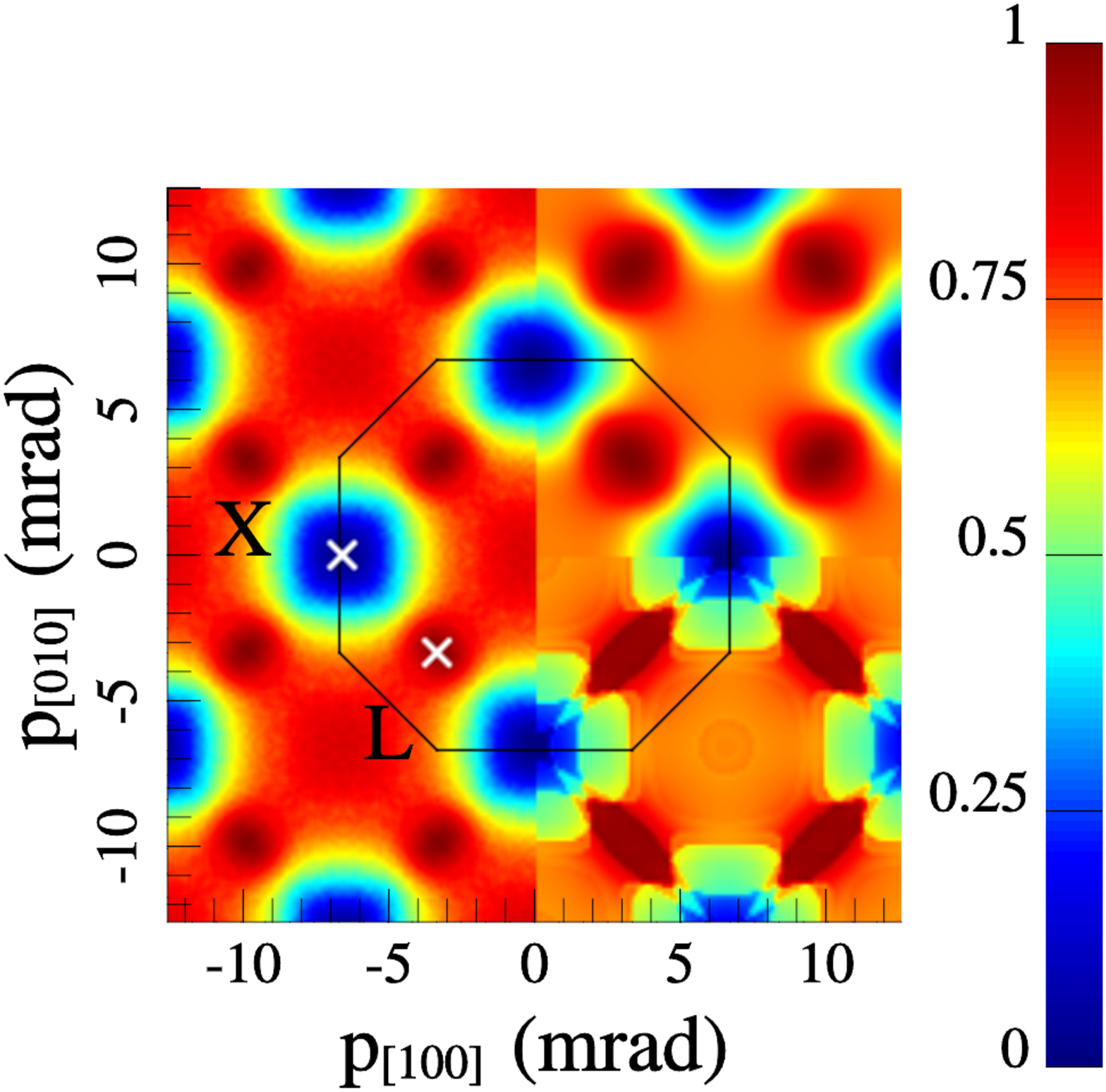}}
\caption{{$[001]$} projection of copper (colour online).}
\end{figure}

In figure \ref{fig:vergleich} the anisotropy of the measured (left) and the calculated data (right) is displayed. Obviously the innermost first order features are in very good agreement: There is an increased density in [110] direction at the L-point due to the necks of the Fermi surface and a decreased density in [100]. The minimum along the latter direction corresponds to the X-point of the Brillouin zone.

This attribute is even more explicit in the LCW folded density on the left side of figure \ref{fig:vergleich2}: The highest density is located at the projected L-points, where the Fermi surface extends to the boundary of the Brillouin zone, whereas the lowest density is again at the X-points. In the case of copper, there are no electronic states at the X-point in the band which crosses the Fermi level and hence a low $\rho^{2\gamma}$ is measured. The right side of figure \ref{fig:vergleich2} displays the theoretical LCW density. For a better comparison, the upper right corner shows the calculation convolved with the experimental resolution. Again, the congruency is very good. The only small difference concerns the size of the necks at the L point.

\begin{figure}[tbp]
\begin{center}
\includegraphics[width=0.35\textwidth]{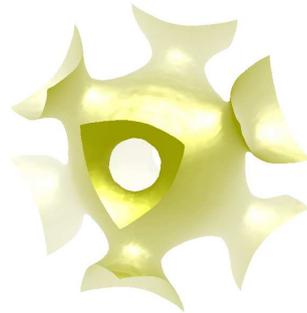}
\end{center}
\caption{\label{fig:fermi} Isosurface plot of the reconstructed 3D $\rho^{2\gamma}$ of copper from only three projections. The density for the isosurface is chosen in such a way, that the resulting isosurface divides the Brillouin zone in two equal parts.}

\end{figure}

To calculate the three dimensional $\rho^{2\gamma}$ a MEA was applied using the three measured projections and all the equivalent directions. The data were then folded into the first Brillouin zone by the LCW procedure for three dimensions. In figure \ref{fig:fermi} the isosurface dividing the Brillouin zone in equal parts is plotted, which has the well known shape of the Fermi surface of copper (see e.g. \cite{PhysRevLett.7.156,PSSB:PSSB2221770210}). Although the qualitative agreement with previous measurements is good, three projections are not sufficient to extract reliable quantitative data. 

\section{Conclusion and Outlook}
We have performed first measurements with the new 2D-ACAR spectrometer at TUM in Garching on the model system copper. The validity of the obtained data could be proven using ab-inito calculations together with LCW folding and MEA. Even though only three projections were recorded, the 3D $\rho^{2\gamma}$ could be obtained.

In the near future, it is planed to measure more projections of copper to compare our results quantitatively. Further investigations will then aim at more complex materials, especially correlated electron systems. Spectra will be collected as a function of temperature in order to investigate phase transitions. In addition we plan to compare the measurements with results obtained by complementary techniques such as, e.g., quantum oscillatory methods.

\section*{Acknowledgements}
 The authors would like to thank the technical staff of the MLL accelerator laboratory and the directorial board, especially L. Beck for his kind support. 
This project is funded by the Deutsche Forschungsgemeinschaft through the Transregio TRR80.

\section*{References}
\bibliography{Paper}

\providecommand{\newblock}{}
\begin{thebibliography}{10}
\expandafter\ifx\csname url\endcsname\relax
  \def\url#1{{\tt #1}}\fi
\expandafter\ifx\csname urlprefix\endcsname\relax\def\urlprefix{URL }\fi
\providecommand{\eprint}[2][]{\url{#2}}

\bibitem{RevModPhys.60.701}
Schultz P~J and Lynn K~G 1988 {\em Rev. Mod. Phys.\/} {\bf 60}(3) 701

\bibitem{Lock1973}
Lock D~G, Crisp V~H~C and West R~N 1973 {\em J Phys.\/} F Met. Phys. {\bf 3}
  561

\bibitem{Gull1978}
Gull S and Daniell G 1978 {\em Nature\/} {\bf 272} 686

\bibitem{0295-5075-32-9-012}
Fretwell H~M {\em et~al.\/} 1995 {\em Europhys. Lett.\/} {\bf 32} 771

\bibitem{springerlink:10.1007/s00339-011-6350-z}
Pylak M, Kontrym-Sznajd G and Dobrzyński L 2011 {\em Appl. Phys.\/} A {\bf
  104}(2) 587

\bibitem{Burch1983113}
Burch S~F, Gull S~F and Skilling J 1983 {\em Comput. Vision Graph.\/} {\bf 23}
  113

\bibitem{Skilling1984}
Skilling J and Bryan R~K 1984 {\em Mon. Not. R. Astron. Soc.\/} {\bf 211} 111

\bibitem{ceeh:043905}
Ceeh H, Weber J~A, Leitner M, B\"{o}ni P and Hugenschmidt C 2013 {\em Rev. Sci.
  Instrum.\/} {\bf 84} 043905

\bibitem{0034-4885-74-9-096501}
Ebert H, K\"{o}dderitzsch D and Min\'{a}r J 2011 {\em Rep. Prog. Phys.\/} {\bf
  74} 096501

\bibitem{PhysRevLett.7.156}
Burdick G~A 1961 {\em Phys. Rev. Lett.\/} {\bf 7}(5) 156

\bibitem{PSSB:PSSB2221770210}
Kondo H {\em et~al.\/} 1993 {\em Phys. Status Solidi\/} B {\bf 177} 345

\end{thebibliography}
\bibliographystyle{iopart-num}


\end{document}